\documentclass[amsmath,nofootinbib,amssymb,twocolumn]{revtex4}
\begin{document}
\title{Distributional Matter Tensors in Relativity}
\author{C. K. Raju}
\affiliation{University of Poona, Pune 411 007, India}
\thanks{This paper was presented at MG5 (1988) and originally published in \textit{Proceedings of the Fifth Marcel Grossmann Meeting on General Relativity}, ed. D. G. Blair and M. J. Buckingham, World Scientific, Singapore, 1989, pp 419--22. Apart from the practical value of the new (and still unused) shock conditions, the paper illustrates an important point in the new philosophy of mathematics proposed in my recent book, \textit{Cultural Foundations of Mathematics}, Pearson Longman, 2007. Hence, it has been reproduced verbatim on the archive (with the addition of two footnotes and the abstract). }

%This abstract not part of original paper.
\begin{abstract}
This paper uses products of distributions to obtain new junction conditions for relativistic shocks. In general, the shock is accompanied by a surface layer, and the new conditions generalize both Taub's jump conditions for shocks, and those of Israel and Kuchar for surface layers. In the non-relativistic limit, the surface layer is present only when the fluid is viscous or thermally conducting---a situation where the classical Rankine-Hugoniot conditions do not apply. Thus, our conditions properly extend all previous conditions, and provide complete Cauchy data needed to solve the full Navier-Stokes equations downstream of the shock. Since the associative law fails for our product, the residual uncertainty regarding the association of factors must be eliminated empirically. This is equivalent to fixing the correct initial \textit{form} of the equations (such as the ``conservation form'' in the Euler case).
\end{abstract}

\maketitle{}

\section{Introduction}

Distributional matter tensors arise in problems concerning shells of matter, shock waves, line sources, and possibly some `singularities' \cite{Israel1979, Taub, GerochTraschen}. Such matter tensors involve problems with products of distributions. Here we give a quick overview of the theory of such products, and point out conditions under which such matter tensors may be regarded as distributional limits of smooth matter tensors.

Distributional matter tensors are believed to be no more than simplified models representing a situation which is actually smooth; however, this belief is not always justified. For example, for perfect fluids, one may arrange initial conditions so that (classically) the characteristics intersect, and genuine discontinuities develop \cite{Richtmeyer}; strings are believed to be represented by 2-dimensional surfaces, and `singularities' might also indicate a breakdown of smoothness.

The geometrical theory of distributions on a manifold is straightforward \cite{Parker} except for the analytical problem of products. According to Taub \cite{Taub}, ``Fortunately the product of such distributions (as arise) is quite tractable.'' Thus, if $\theta$ is the Heaviside function, $\theta^2 = \theta$ would suggest $2 \theta \cdot \theta ^\prime = \theta ^\prime$, i.e., $\theta \cdot \delta = \frac{1}{2}\delta$. This is simple enough, except that $\theta^3 = \theta$ would suggest $3 \theta^2 \cdot \theta ^\prime = \theta ^\prime$ which should be the same as $\theta \cdot \delta = \frac{1}{3}\delta$. More generally, since $x^{-1} (x \delta )= 0 \neq \delta = (x^{-1} x ) \delta$, the Schwartz impossibility theorem asserts that there does not exist an associative differential algebra $A \supseteq D^\prime$ such that the product in A agrees with the Schwartz product or the product of $C^\infty$ functions. 

In contrast to Taub's naive approach, Parker's suggestion to use H\"ormander's product would exclude even the theory of shells, since $\theta \cdot \delta$ is not defined then. At the other extreme, Geroch and Traschen have proposed a notion of a regular metric (including possibly discontinuous metrics) in such a way as to exclude entirely products of distributions. This seems too strict, since such products have been freely used in renormalization theory and are known to `work'! Such regularity conditions also depend on the precise form of the equations used, since p.d.e.'s with equivalent smooth solutions may have inequivalent weak solutions. 

\section{Products of distributions}

Currently, the real problem is to select one definition from amongst the many that are available. According to the Fourier transform method \cite{Hormander, Ambrose, ReedSimon, Vladimirov}, the product is essentially defined for $f, ~g \in D^\prime$ by $ f \cdot g = ( \hat{f}  \otimes \hat {g} )^{ \check{}} $, whenever the r.h.s. exists, $\otimes$ being convolution. This product is \cite{Tysk, Oberguggenberger} a particular case of the sequential product \cite{Mikusinski, HirataOgata, Fisher, Kaminski} $f \cdot g = D^\prime-\lim (f \otimes \delta_n) \cdot ( g \otimes \rho_m)$ whenever the r.h.s. exists for `all' $\delta$-sequences $\delta_n$, $\rho_m$. Putting 
$\delta_n \equiv \rho_m$ = an approximate identity, and requiring the limit to be independent of the choice, we get the sequential `model' product \cite{OberguggenColomb}. Tighter conditions on $\delta_ n$ would allow more products to exist, but could lead to a situation where physically acceptable regularisations have non-unique limits. 

Colombeau's \cite{ColombeauNGF, ColombeauEINGF} associative differential algebra $G \supseteq D^\prime$, and the product does not agree with the product of $C^\infty$ functions. The product of any $f,~g \in D^\prime$always exists in $G$ but admits an `associated' distribution iff \cite{Jelinek} the sequential model product exists. However, the product is not coherent with association so that $\delta \cdot \delta $ need not have a unique associated distribution. This product is thus analogous \cite{Todorov} to the product in the non-standard space $^*E = ^*C^\infty$. A similar ambiguity exists for products defined by the Hahn-Banach method used in renormalization theory. There one defines \cite{Jaeger, DurandBremermann} $\partial ^ \alpha ( \hat f \otimes \hat g) = \hat f \otimes \partial ^ \alpha \hat g$, provided the r.h.s. exists. The Hahn-Banach extension to the whole space is non-unique upto arbitrary linear combinations of $\delta$ and its derivatives to order $\alpha$. Thus $\theta \cdot \delta = A \delta$, $\delta ^ 2 = B \delta$, with $A$, $B$ arbitrary constants. 

I define \cite{ckr82} $ f \cdot g = \frac{1}{2}  ~^*\left\{( f \otimes \delta _\omega ) g + (g \otimes \delta _ \omega ) f \right\}$, where * denotes the non-standard extension to $^*D ^ \prime$ (i.e., $^*E$) and $\omega$ is a positive infinite integer. The Leibniz rule holds, but the associative law fails. The product of any $f, ~ g \in D ^ \prime$ always exists, is unique and coincides with the sequential product when the latter exists. We have $\theta \cdot \delta = \frac{1}{2} \delta$, and $\delta^2 = \delta_\omega (0) \delta$, where $\delta _ \omega (0 )$ is infinite. Nevertheless, for applications to weak solutions of p.d.e.'s a kind of linear independence \cite{ckr82, ckr82b} ensures that the final results are standard. For applications to renormalization, it may be shown that the infinities arising in this way actually do cancel \cite{ckr83}.

\section {Shock waves in real continua}

For reasons of space and time, I will mention here only applications to shock waves in real continua (e.g. in fluids with viscosity). The problem here is \textit{not} that of using a smooth solution of the viscous-fluid equations to interpolate a shock solution in a perfect fluid (as e.g. in Ref. 4). Rather the problem is to develop solutions of the viscous equations into the region downstream of the shock. For this purpose the classical conditions of Taub or Rankine-Hugoniot fail to provide adequate Cauchy data on the shock hypersurface $\Sigma$, and the standard interpolation techniques are also useless. 

For timelike $\Sigma$, the jump conditions across $\Sigma$ are
\begin{align}
	[Nu^1] = 0,  \tag {3.1} \label{one}\\
	\tilde{T}^{\alpha 1} = 0, \label {two} \tag{3.2}\\ 
	\tilde {T}_{| j} = - [T ^{i1}], \label{three} \tag{3.3}\\ 
	K_{ij} | \tilde{T}^{ij} = -[T^{11}]. \label{four} \tag{3.4}
\end{align}

\noindent Here $\tilde T = \lim <T, ~\phi>$ as $\rm{supp}~ \phi \downarrow \Sigma$,
for $\phi \in D$ with $\phi = 1$ on a neighbourhood of $\Sigma$, $T$ being the stress tensor of the continuum, $_{|j}$ denotes covariant differentiation w.r.t. the intrinsic geometry of $\Sigma$, [] denotes the jump, and $|$ the mean value. (I am not convinced that the stress tensor can be defined without using admissible coordinates: $x^1$ is the gaussian coordinate normal to $\Sigma$.)

Note that a surface layer is present, and \eqref{two} -- \eqref{four} are the conditions for surface layers except that \eqref{two} is identically satisfied in the formalism of Israel \cite{israelN, IsraelNb} and Kuchar \cite{Kuchar}. For a perfect fluid, all terms with a \~{} vanish and the conditions reduce to the Taub conditions. But for a viscous and thermally conducting fluid, the nonrelativistic limit is 

\begin{gather}
	[\rho v] = 0, \label {nr-one}\tag{3.1$^\prime$}\\ 
	\tilde{\rho} = (2 \eta + \lambda) [v],	\quad \tilde{e} = \frac{\kappa}{v |} [T], \label {nr-two} \tag{3.2$^\prime$}\\
   \left [ p + \rho {v^2} \right ] = (2 \eta + \lambda) [ v , _{ n} ], \label {nr-three} \tag{3.3$^\prime$}\\
	(\rho v) | [w + \frac{1}{2} v^2] = (2 \eta + \lambda) [v v , _{n}] + \kappa [T , _{n} ] - \tilde {e} , _{t}, \label{nr-four} \tag{3.4$^\prime$}
\end{gather}

\noindent where $v$ is the fluid velocity in the direction of the normal $n$ to the shock, terms with a \~{} are defined as before, $w$ is the enthalpy, and the remaining notations are standard. Thus \eqref{two} is non-trivial. 

\eqref{one}--\eqref{four} are still insufficient to determine adequate Cauchy data on $\Sigma$, since an arbitrary equation of state may not be used on $\Sigma$, but, using the first law of thermodynamics, and $\delta ^ 2$, one gets
\begin{gather}
	[\mu] = \frac{1}{N|} [p] + T | h_1 [S],\label {five} \tag{3.5}\\
	\tilde{\rho} = \frac{(NT) [N]}{[NT]} \tilde{\mu} + \frac{h_0}{v|},\label{six} \tag{3.6}
\end{gather}

\noindent where $h_0$ and $h_1$ are uncertainty factors due to the failure of the associative law,\footnote{This is a desirable feature of the theory, since the failure of the associative law explains why p.d.e.'s with equivalent smooth solutions may have inequivalent discontinuous solutions---a counter-intuitive point on which Riemann slipped (he incorrectly supposed that energy and entropy conservation were equivalent across a shock). This simple explanation is not available with associative products like Colombeaus's, and  an earlier attempt (Raju and Colombeau 1987, unpublished) to compare this product with Colombeau's by checking possible differences in the resulting junction conditions (for the non-relativistic case) was abandoned.} $S$ is the entropy, and $\mu$ is the chemical potential. \eqref{one}--\eqref{six} suffice to fix Cauchy data on $\Sigma$. Incidentally, the weak-shock non-relativistic limit of \eqref{five} is

\begin{equation}
	T|[S] = \frac{1}{4V| j^4} [p]^3 + (2 \eta + \lambda ) v , _{ n} | [v] + \frac{\kappa}{j} [T , _{ n}] - \frac{1}{j} \tilde {e} , _{t}, \label {nr-five} \tag{3.5$^\prime$}
\end{equation}

\noindent giving an exact expression for the entropy change.

The distributional matter tensors may also be thought of as simplified models: the weak solutions $g_{\alpha \beta} \in C ^0$ are the uniform limit on compacta of the smooth $g_{\alpha \beta} \otimes \delta _ n$ which satisfy Einstein's equations with the matter tensor
\begin{equation}
	8 \pi (T^\prime _ {\mu \nu})_n = 8 \pi T_{\mu\nu} \otimes \delta_n ) - G _{\mu \nu} \otimes \delta_n. \label {seven}\tag{3.7}
\end{equation}

\noindent $(T^\prime _ {\mu \nu})_n \rightarrow T_{\mu\nu}$ in $D^\prime$ since $G^\prime _ {\mu \nu} \rightarrow G_{\mu\nu}$ in \eqref{seven}, provided only that the products of distributions appearing in Einstein's equations exist according to the sequential model product. A background of positive energy is a sufficient condition for positivity of energy with this $T^\prime _ {\mu \nu}$. If associativity is taken for granted, and products such as $\theta \cdot \theta \cdot\delta$ arise, different regularisations would have different limits. On the other hand, it may be possible to make sense out of a wider class of genuine line sources if my product is used, and a definite association of factors can be fixed. 
\begin{acknowledgments}
The author gratefully acknowledges financial assistance from ICTP which enabled him to attend the MG5.
\end{acknowledgments}

\bibliography{proddist}

\end{document}